\documentclass[preprint,showpacs,preprintnumbers]{revtex4}

\usepackage{graphicx}
\usepackage{dcolumn}
\usepackage{bm}
\usepackage{epsfig}
\usepackage{epstopdf}
\usepackage{grffile}
\usepackage{color}
\usepackage{xcolor}
\usepackage{colordvi}
\usepackage{array}
\usepackage{tikz} 
\usepackage{tikz-feynman}
\usepackage{amsmath}
\usepackage{amssymb}
\usepackage{rotating}
\usepackage{lscape}
\usepackage{float}
\usepackage{footnote}
\usepackage{wasysym}
\usepackage{gensymb}
\usepackage{subcaption}
 \makesavenoteenv{environmentname}

\usepackage{hyperref}

\begin{document}


\title{Dark matter interpretation of GRB 221009A: a singlet scalar explains LHAASO data}

\author{S. Peyman Zakeri}
\email{peyman.zakeri@ipm.ir}
\affiliation{School of Physics, Institute for Research in Fundamental Sciences (IPM), P.O.Box 19395-5531, Tehran, Iran}

\date{\today}

\begin{abstract}
In this work, we propose a dark matter (DM) interpretation of the intensive gamma ray burst GRB 221009A. This indirect detection approach devotes to the decay of DM particles into high energy (HE) photons. In this context, a singlet scalar DM generated at the redshift of GRB 221009A is up-scattered by the high energy cosmic rays (HECRs) during its propagation to Earth. This highly boosted DM then possesses a high flux and undergoes a dominant di-photon decay before reaching the detector. The Large High Altitude Air Shower Observatory (LHAASO) probes such energetic gamma rays whereby has recorded a $\sim$13 TeV event for the aforementioned GRB.

\vspace{0.3cm}
\end{abstract}

\maketitle


\section{Introduction}\label{sec:int} 
GRBs are considered as astrophysical sources to generate HECRs. These phenomena manifest as instant HE explosions that release transient jets dominantly composed of photons and electron-positron pairs. In this regard, HE photons have drawn considerable attention from astroparticle physics and cosmology. Following that, LHAASO is a renown project seeking for HE gamma ray using electron and muon detectors \cite{Cao:2010zz}. As well as that, the Swift-Burst Alert Telescope \cite{2022GCN.32635....1K}, the Fermi satellite \cite{2022GCN.32636....1V} and Carpet-2 experiment \cite{2022ATel15669....1D}, which reported 251 TeV photon-like air shower, were also associated with the GRB 221009A. Recent LHAASO results indicate that the most reliable photon energies observed from GRB 221009A are in the range of 12–13 TeV \cite{LHAASA:2023pay}, although initial reports suggested values up to 18 TeV \cite{2022GCN.32677....1H}. However, it is important to note that such TeV-scale photons are unable to reach Earth due to interactions with extragalactic background light (EBL), resulting in moderate attenuation \cite{Gould:1966pza, Fazio:1970pr}. Thus, the puzzle surrounding the detection of HE photons from GRB 221009A is not solely related to their energy, but rather to their survival as they propagate through the intergalactic medium. While standard explanations for this observation exist, such as intrinsic transparency or inverse Compton scenarios \cite{Sahu:2022gvx, Sahu:2024ddo}, these do not fully address the photon attenuation expected from EBL. Non-standard mechanisms, including Lorentz invariance violation \cite{Li:2022wxc, Finke:2022swf} have also been proposed. Besides, Numerous theoretical proposals have emerged, suggesting various candidates such as axion-like particle (ALP) \cite{Baktash:2022gnf, Galanti:2022chk, Lin:2022ocj, Troitsky:2022xso, Nakagawa:2022wwm, Zhang:2022zbm, BetancourtKamenetskaia:2024gcv}, neutrino \cite{Stecker:2016fsg, Smirnov:2022suv, Brdar:2022rhc, Huang:2022udc}, and scalar \cite{Balaji:2023nbn, Borah:2023sal} that may facilitate the transport of these photons. These fields are embedded beyond the standard model (BSM), allowing for interactions with photons through mediators.

One of the cosmological puzzles that can potentially be addressed through BSM physics is DM. In this context, new fields can play the role of DM, with their interactions with bath particles being characterized by appropriate mediators \cite{Gondolo:1990dk, Srednicki:1988ce, Silveira:1985rk, McDonald:1993ex, Burgess:2000yq, Barger:2007im, Guo:2010hq, Yaguna:2011qn, Cline:2013gha, Rezaei:2020mqj, McDonald:2001vt, Hall:2009bx, Hambye:2008bq, Hambye:2009fg, YaserAyazi:2015egj, YaserAyazi:2018pea, PeymanZakeri:2018zaa, Babu:2014pxa, Lebedev:2021xey, He:2024hkr}. The Higgs portal is one of the eminent approach in which the Higgs boson mediates the interaction between DM particles and those of the standard model (SM) \cite{Lebedev:2021xey}. Within this framework, a singlet scalar can yield a minimal DM model, characterized by a reduced number of independent free parameters \cite{Silveira:1985rk}. Taking phenomenological aspects, there are a wide variety of experiments in order to investigate both direct and indirect evidence of DM. Due to the lack of direct signal in terrestrial experiments such as XENONnT \cite{XENON:2023cxc} and LUX-ZEPLIN (LZ) \cite{LZ:2022lsv}, scientists have increasingly focused on indirect detection methods. This approach considers the products of DM annihilation and decay, including particles such as electrons, positrons, protons, antiprotons, photons, and neutrinos. Notably, photons have been successfully detected in various experimental setups.

In recent years, accelerating DM by various celestial boosters into relativistic velocities and ultra-high energies (UHE) has been considered \cite{Agashe:2014yua, Bhattacharya:2014yha, An:2017ojc, Emken:2017hnp, Ema:2018bih, Bringmann:2018cvk, Wang:2019jtk, Cappiello:2019qsw, Guo:2020drq, Wang:2021jic, Xia:2021vbz, Jho:2021rmn, Chao:2021orr, Granelli:2022ysi, Carenza:2022som, DeRomeri:2023ytt, Wang:2023wrx, Das:2024ghw, Kumar:2023nqi, Huang:2025xwz}. This phenomenon represents a promising avenue for direct detection efforts. Specifically, light DM as well as candidates following freeze-in mechanism (with feeble interaction), can leave measurable nuclear recoils in underground detectors. Furthermore, boosted DM can attain sufficient flux to generate HE particles through annihilation and decay processes. Consequently, boosting DM may provide solutions to existing cosmic anomalies.       

In this letter, we propose a hypothesis that a well-motivated DM candidate, i.e. singlet scalar, can explain LHAASO puzzling observation of GRB 221009A. Two scenarios for the production of this DM are considered: either as relic DM generated in the early Universe or via Higgs decay at a redshift of $z=0.1505$, possibly facilitated by HE interactions near the GRB. Subsequently, this DM candidate achieves relativistic velocity through interactions with protons. The HE DM then decays upon reaching Earth, resulting in the emission of photons with energy of about 13 TeV, which can be detected by LHAASO. The focus on GRB 221009A stems from its extraordinary energy output and relatively close redshift ($z=0.1505$). While similar mechanisms might exist in other astrophysical events, GRB 221009A's rarity and alignment with the LHAASO detection of $\sim$13 TeV photons make it a uniquely compelling case for probing DM-related phenomena. 

The rest of the paper is organized as follows. Section \ref{sec:SDM} provides the structure of singlet scalar DM model. Then, the process of boosting DM through the extragalactic HECRs is presented in section \ref{sec:BST}. In section \ref{sec:Flx}, we compute the photon flux resulted from the scalar DM decay at Earth. In section \ref{sec:Res}, we present our findings about imprinting of DM in LHAASO data. Finally, we discuss the implications and concluding remarks in the last section.

\section{Singlet scalar DM}\label{sec:SDM}

\subsection{The Model}

In the sequential model, DM is capable of interacting with the thermal soup via the Higgs portal, with its stability typically ensured by a discrete $Z_{2}$ symmetry. The extreme conditions surrounding GRB 221009A make it a suitable environment for the production of Higgs bosons, which subsequently decay into DM particles (Higgs $\rightarrow$ DMDM). While DM production could theoretically occur at other redshifts, the unique observational data from GRB 221009A enable us to propose a specific, testable hypothesis linking DM production to the astrophysical properties of this event. The parameters of this model are defined as follows
\begin{equation}
\label{eq1}
\mathcal{L} = \mathcal{L}_{SM} + \frac{1}{2} \partial_{\mu} S \partial^{\mu} S - \frac{m_{0}^{2}}{2} S^{2} - 
\frac{\lambda_{S}}{4} S^{4} - \lambda_{HS} S^{2} H^{\dagger} H.
\end{equation}
In this literature, $S$ represents the DM, $m_0$ its bare mass and $H$ signifies the $SU(2)$ Higgs doublet. Following the process of spontaneous symmetry breaking, we have
\begin{equation}
H = \frac{1}{\sqrt{2}} \left( \begin{array}{c}0  \\v_{H}+h\end{array} \right)\,,
\end{equation}
where $v_H=246$ GeV represents the vacuum expectation value of the Higgs field, leading to a modification of the Lagrangian
\begin{equation}\mathcal{L} = \mathcal{L}_{SM} + \frac{1}{2} \partial_{\mu} S \partial^{\mu} S - \frac{1}{2} m_S^2 S^{2} - 
\frac{\lambda_{S}}{4} S^{4} - \lambda_{HS} S^{2} h^{\dagger}h-\lambda_{HS}\,\,v_H S^{2}h.
\end{equation}
Here, DM mass is $m_S=\sqrt{m_0^2+\lambda_{HS}v_H^2}$ and $h$ is the SM-like Higgs detected at the LHC. The Higgs boson serves as the sole mediator linking the singlet scalar DM to the SM particles. Consequently, all final states are influenced by the Higgs-DM coupling, which is characterized by \( \lambda_{HS} \). Furthermore, the $H-S$ mixing angle is also \cite{Dev:2020jkh}
\begin{equation}
\label{eq5}
\sin\theta_{HS} \simeq \frac{\sqrt{2}\lambda_{HS} v_{H}v_{S} }{m_h^2},
\end{equation}
which is an essential parameter in our research. Now, we evaluate the phenomenological aspects of the scalar DM. 
\subsection{Relic Density}
In examining the relic abundance of DM, different mechanisms through which $S$ was produced in the early Universe should be considered. In this work, the singlet scalar $S$ can have two possible origins. First, it can be generated via Higgs boson decay in the vicinity of the GRB 221009A, where extremely high temperatures and particle densities may facilitate Higgs production through proton-proton or photon-photon collisions. If the Higgs boson of the SM is primarily responsible for DM production, relic density of $S$ will be obtained in terms of mixed $H-S$ coupling $\lambda_{HS}$ as \cite{Babu:2014pxa}  
\begin{equation}
	\Omega_S h^2\simeq0.12 (\frac{\lambda_{HS}}{4.5\times 10^{-9}})^{2},
\end{equation}
which is experimentally measured as $\Omega_{\text{DM}}h^2=0.120\pm0.001$ \cite{Planck:2018vyg}. Thus, the parameter space considered in the following sections should be compatible with the relic density constraint in this scenario. Alternatively, $S$ can also originate from the relic background DM distributed across the universe. In our calculation of the photon flux arriving at Earth, we adopt this relic DM scenario, where $S$ is assumed to exist as a background field and is boosted via scattering with HECRs emitted by GRB 221009A.

\subsection{Direct Detection}
Direct search strategies for DM involve significant contributions from nuclear recoil effects. In this context, the scattering of DM with nucleons generates detectable signals in terrestrial experiments. The cross section for DM-nucleon interactions, which arises from the coupling between DM and SM particles, is parametrized as \cite{He:2024hkr}
\begin{equation}
	\label{eq6}
	\sigma_N \simeq \frac{4\alpha q_N^2 \mu_N^2}{\Lambda^4} .
\end{equation}
Here, $\alpha$ is the fine structure constant, $q_N(A, Z)$ is the DM-nucleus effective coupling, $\mu_N=m_Nm_S/(m_N+m_S)$ denotes the reduced mass and $\Lambda=M(e/|c|)^{\frac{1}{2}}$ is interpreted as the effective cutoff scale. The spin-independent scattering cross section derived from this analysis should be juxtaposed with the experimental data obtained from XENONnT \cite{XENON:2023cxc} and LUX-ZEPLIN (LZ) \cite{LZ:2022lsv} in order to explore the parameter space associated with the scalar DM. Passing from direct search, we focus on strategies that investigate DM through indirect means.

\subsection{Indirect Detection: Photon Production from DM Decay}
Now, we consider radiative decay of such a DM wherein HE photons are generated. Although $S$ can decay to other secondary particles (charged leptons for instance), the dominant decay mode here is $S\rightarrow\gamma\gamma$, provided that $m_S<2m_e$ (the lightest lepton). The decay rate of this process is  
\begin{equation}
	\Gamma(S\rightarrow\gamma\gamma)=(\frac{\alpha}{4\pi})^2|F|^2\sin^2\theta_{HS}\frac{G_Fm_S^3}{8\sqrt{2}\pi},
\end{equation}
where $G_F$ represents Fermi constant, form factor $F$ is
\begin{equation}
	F=F_W(\beta_W)+\sum_f N_C Q_f^2 F_f(\beta_f),
\end{equation}
and
\begin{equation}
	\beta_W=\frac{4m_W^2}{m_S^2},~~~~~~~~\beta_f=\frac{4m_f^2}{m_S^2}. 
\end{equation}
$F_W$ and $F_f$ (in the summation) are given as
\begin{equation}
	F_W(\beta)=2+3\beta+3\beta(2-\beta)f(\beta),
\end{equation}
\begin{equation}
	F_f(\beta)=-2\beta[1+(1-\beta)f(\beta)],
\end{equation}
with
\begin{equation}
	f(\beta)=\arcsin^2[\beta^{-1/2}].
\end{equation}
According to the approximation utilized in \cite{Babu:2014pxa}, $F_W=7$, $F_f=-4/3$ and hence $F=-11/3$. It should be noted that $Q_f$ and $N_c$ denote the electric charge and the number of color charge respectively. Ultimately, our DM contributes to the photon flux with the following branching ratio
\begin{equation}
	B_\gamma\equiv\text{Br}(S\rightarrow\gamma\gamma)=\frac{\Gamma(S\rightarrow\gamma\gamma)}{\Gamma_{S}},
\end{equation}
where $\Gamma_{S}$ is the DM decay rate and $\Gamma(S\rightarrow\gamma\gamma)$ corresponds to the decay channel $S\rightarrow\gamma\gamma$. 
\section{Boosted DM}\label{sec:BST} 
In this section, we explore the possibility that HECRs can significantly up-scatter DM particles during their journey from redshift $z=0.1505$ to $z=0$ (Earth). The GRB 221009A plays a critical role in both the acceleration and potential production mechanisms of DM.
In the relic DM scenario, the extreme energetic environment near GRBs can facilitate processes such as Higgs boson production, potentially leading to DM generation.  In addition, GRB 221009A serves as a powerful source of HECRs, which interact with the background DM, boosting it to high energies. Thus, the GRB indirectly contributes to the mechanism by supplying the HECRs required for up-scattering. 

While protons originating from the GRB contribute to the acceleration of DM particles, they alone are insufficient to account for the observed high flux and energy of DM. Instead, our model considers the cumulative interactions of DM particles with cosmic ray protons along the entire path from the GRB location to Earth. This broader interaction framework ensures that DM particles gain sufficient energy and flux to produce the observed HE gamma rays upon decay. 

In this manner, we assume that cosmic boosters are composite of nucleons purely and take protons and neutrons in the same way (it is supposed that they have equal cross-section with DM). Furthermore, following \cite{Berezinsky:2002nc}, we consider energy range of protons as $E_p\lesssim 10^{18}$ eV in order to evade from the attenuation in the path to Earth. Nonetheless, the spectral features are not explicit below this EeV-scale, we utilize ultra-high energy cosmic rays (UHECRs) and then extrapolate it to our interested regime \cite{Guo:2020drq}. Even though there are several proposals elucidating the spectral features of UHECRs, we adopt the dip-model \cite{Berezinsky:2002nc} in this work. This framework is based on the interaction of UHE (extragalactic) protons with cosmic microwave background (CMB) using four features of the proton spectrum: GZK cutoff, bump, dip and the second dip. In addition, we assume that DM is at rest and ignore its initial negligible velocity. Now, the flux of UHECR boosted DM is 
\begin{equation}\label{Eq.14}
	\Phi_S=\int\phi_S d\Omega=4\int_{b_{\text{min}}}^{b_{\text{max}}}db\cos b\int_{l_{\text{min}}}^{l_{\text{max}}}dl\phi_S.
\end{equation}
Here, we have integrated over solid angle as GRB221009A is a transient jet whose energy is released isotropically into a spherical fireball. $l$ and $b$ are galactic longitude and latitude respectively and $\phi_S$ is the differential DM flux as follows \cite{Guo:2020drq}
\begin{equation}\label{Eq.17}
	\phi_S=\int_{0}^{E_{p\text{max}}}dE_p\int_{0}^{0.1505}dz\sigma_{pS}\bar{n}_S(1+z){\cal F}(E_p,z)D_{pS}(E_p,E_S(1+0.1505))|\frac{cdt}{dz}|,
\end{equation}
where $\sigma_{pS}$ is the DM-proton scattering cross section. Besides, the corresponding average DM number density at redshift $z=0.1505$ is estimated to be 
\begin{equation}
	\bar{n}_S=\bar{n}_S^0(1+z)^3|_{z=0.1505},
\end{equation} 
with $\bar{n}_S^0\simeq 10^{-6}(\frac{\text{GeV}}{m_S})\text{cm}^{-3}$ \cite{Masaki:2011yc} the current DM number density at $z=0$. The term $(1+z)$ in the integral is considered when the energy scale is reduced due to the expansion of the Universe \cite{Guo:2020drq}. The function ${\cal F}(E_p,z)$ represents diffuse cosmic ray (CR) flux and for protons of energy $E_p\lesssim 10^{18}$ eV, can be written as \cite{Berezinsky:2002nc}
\begin{equation}
	{\cal F}(E_p,z)=\frac{c}{4\pi}\int_{z}^{\infty}{\cal S}(z')\frac{1+z'}{1+z}F_{\text{inj}}(\frac{1+z'}{1+z}E_p)|\frac{dt}{dz'}|dz',
\end{equation}
where the factor of ${\cal S}(z)$ is attributed to the historical evolution of the cosmos and $F_{\text{inj}}(E_p)$ accounts for the spectrum of injected proton as below \cite{Guo:2020drq}: 
\begin{eqnarray} \label{Eq:7}
	F_{\text{inj}}(E_p) = N_0
	\left\{
	\begin{array}{ll}
		E_c^{-\gamma+2}E_p^{-2},~~~~  & \mbox{if } E_P < E_c, \\
		E_p^{-\gamma},~~~~ & \mbox{if } E_p \geq E_c,
	\end{array}
	\right.
\end{eqnarray}
which is normalized to the observed UHECR spectrum by $N_0$, with $E_c$ set at $10^{15}$ eV \cite{Guo:2020drq}. The parameter $\gamma$, representing the spectral index, will be determined in the subsequent analysis. Numerous injection models exist that incorporate appropriate values for ${\cal S}(z)$ and $\gamma$ \cite{AlvesBatista:2019rhs}. Among the three cosmological models discussed in the \cite{Guo:2020drq}, we select model 2, which specifies $\gamma=2.5$ and
\begin{eqnarray}
	  {\cal S}(z)=
	   \left\{
	\begin{array}{ll}
		(1+z)^{3.4},~~~~ & z\leq 1 \\
		2^{3.7}(1+z)^{-0.3},~~~~ & 1 < z \leq 4 \\
		2^{3.7}\times 5^{3.2}(1+z)^{-3.5}.~~~~ & z > 4 \\
	\end{array} 
	\right. 
\end{eqnarray}
In this specified model, the DM flux has the maximum value, as elaborated in Section \ref{sec:Res}. The next factor in Eq. \ref{Eq.17}, $D_{pS}(E_p,E_S)$ is a transfer function that encodes the energy spectrum of particle DM boosted by a scattering with proton. Provided that the aforementioned scattering is elastic, $D_{pS}(E_p,E_S)$ can be written as
\begin{equation}
	D_{pS}(E_p,E_S)=\frac{\Theta[E_S^{\text{max}}-E_S]}{E_S^{\text{max}}(E_p)},
\end{equation}
where for $E_p\gg m_p, m_S$, we have
\begin{equation}
	E_S^{\text{max}}\backsimeq\frac{E_p}{1+\frac{m_p^2+m_S^2}{2m_SE_p}}.
\end{equation}
The last factor is the Hubble function and is given by 
\begin{equation}
	|\frac{dt}{dz}|^{-1}=H_0(1+z)\sqrt{\Omega_\Lambda+\Omega_M(1+z)^3},
\end{equation}
which is calculable within the $\Lambda$CDM model where $H_0=70$ $\text{kms}^{-1}\text{Mpc}^{-1}$, $\Omega_\Lambda=0.7$ and $\Omega_M=0.3$.   
\section{Photon flux at earth}\label{sec:Flx}

\subsection{Decay Probability and Photon Production}

In this section, we revisit the gamma rays emission by DM decay ($S\rightarrow\gamma\gamma$) at Earth. First, we compute the decay probability of this phenomenon within the distance interval $[x,x+dx]$ \cite{Smirnov:2022suv}  
\begin{equation}\label{Eq.23}
	P_{\text{decay}}=B_\gamma e^{-x/\lambda_{S}}\frac{dx}{\lambda_{S}}e^{-(d-x)/\lambda_{S}},
\end{equation}
where
\begin{equation}\label{Eq.24}
	\tau=\frac{d}{\lambda_{\gamma}},~~~~\lambda_{S}=\frac{E_S}{m_S \Gamma_{S}}. 
\end{equation}
The mean free path of DM is denoted as $\lambda_{S}$, while $\tau$ represents the photon absorption length and is determined in \cite{Gao:2023csq}. Integrating this probability over $x$ and multiplying by DM flux (Eq. \ref{Eq.14}), we can derive the gamma flux
\begin{equation}\label{Eq.25}
	\Phi_\gamma=\frac{2\Phi_S B_\gamma}{\lambda_{S}/\lambda_{\gamma}-1}(e^{-d/\lambda_{S}}-e^{-d/\lambda_{\gamma}}).
\end{equation}
The inclusion of a factor of 2 is necessary to represent the generation of two photons during the decay process of $S$. Now, using $\tau$ definition in Eq. \ref{Eq.24} leads to
\begin{equation}\label{Eq.26}
	\Phi_\gamma=\frac{2\Phi_S B_\gamma}{\tau\lambda_{S}/d-1}(e^{-d/\lambda_{S}}-e^{-\tau}),
\end{equation}
which behaves as a function of photon absorption length.

\subsection{Arrival Time Delay Analysis}
In this section, we estimate the expected arrival time delay between the photons produced by the boosted DM decay and the prompt GRB photons. Since the scalar DM has a non-zero mass, its velocity $v_{\text{DM}}$ after up-scattering will be slightly less than the speed of light $c$. The time delay $\Delta t$ between the DM-induced photon and the prompt GRB photon, over a distance $D$ is approximately

\begin{equation}
\Delta t \simeq D \left( \frac{1}{v_{\text{DM}}} - \frac{1}{c} \right).
\end{equation}
Given the high kinetic energy $E_S$ acquired during up-scattering, we have
\begin{equation}
v_{\text{DM}} = c \sqrt{1 - \left( \frac{m_S}{E_S} \right)^2 }.
\end{equation}
For a scalar DM mass $m_S<2m_e$ $(\cal{O}$(MeV)) and a boosted energy $E_S>2E_\gamma$ $(\cal{O}$(TeV)), the velocity is extremely close to $c$ leading to a negligible time delay. A rough estimation yields
\begin{equation}
\Delta t\lesssim \text{a few seconds}
\end{equation}
which falls within the uncertainty of GRB temporal measurements (typically on the order of seconds to minutes).

We conclude that the DM-induced photons can be considered temporally coincident with the prompt GRB emission. In addition, astrophysical uncertainties such as dispersion in DM velocities and propagation paths introduce only minor corrections to the arrival time, and do not significantly affect the timing coincidence with GRB 221009A.
In the subsequent section, we perform a numerical analysis to elucidate the role of DM in the observations made by LHAASO.

\section{GRB221009A gamma rays: Results and Disscusion}\label{sec:Res} 
In this section, we provide the results of the scenario we have introduced to evaluate its effectiveness in elucidating GRB 221009A. First, it should be noted that one of the major challenges for DM models is providing the accurate relic density in the Universe. Applying this astrophysical constraint to our light DM (of mass $\cal{O}$(MeV)), impose very weak coupling between the dark sector and the SM one \cite{Dev:2020jkh}. More precisely, DM-Higgs coupling $\lambda_{HS}\simeq10^{-9}$ or equivalently mixing angle $\sin\theta_{HS}\simeq10^{-16}$ is required (see Fig. 5 of \cite{Dev:2020jkh}). Besides, direct detection experiments like  XENONnT \cite{XENON:2023cxc} and LUX-ZEPLIN (LZ) \cite{LZ:2022lsv} struggle to place meaningful bounds. However, the sensitivity of current experiments drops off significantly for low-mass candidates (e.g., MeV scale). In this sense, our DM can evade this constraint due to the combination of its small mass and weak interaction. 

In addressing the photon production resulting from DM decay at Earth, it is important to recognize that the previously mentioned selection of the mass-mixing parameter imposes a strict constraint on the DM flux $\Phi_S$. Considering number of produced photons in DM decay, $N_\gamma^S$, we can extract the following relationship for $\Phi_S$ \cite{Balaji:2023nbn}       
 \begin{equation}\label{Eq.27}
 	\Phi_S\simeq(\frac{m_S^2}{1.57\times 10^{-26} \text{GeV}})^{-2}\frac{N_\gamma^S}{\sin^2\theta_{HS}},
 \end{equation}
where we express $N_\gamma^S$ as the number of events given by
\begin{equation}
	N_\gamma^S=\Delta t \int\text{Area}\times\Phi_\gamma dE_{\gamma}.
\end{equation}
Here, $\Delta t$ is the observation time and the integration is over the $\gamma$ energy from effective area of detector multiplying $\gamma$ flux. Using the instrumental features of LHAASO in the energy range $E_{\gamma}\sim (0.5-13)$ TeV leads to three benchmarks for the number of events $N_\gamma^S=50,500$ and 5000. This analysis, thus, results in large DM flux of value $\Phi_S\simeq10^3, 10^4$ and $10^5$ GeV$^{-1}$ cm$^{-2}$ s$^{-1}$ for the aforementioned $N_\gamma^S$s respectively. Subsequently, we need a very extreme astrophysical environment to provide very high density and energy conditions to produce the highly boosted DM. This is why we have taken into account the contributions of all extragalactic cosmic protons (from GRB location to Earth) as diffuse sources. 

\begin{figure}[H]
\centerline{\includegraphics[width=.60\textwidth]{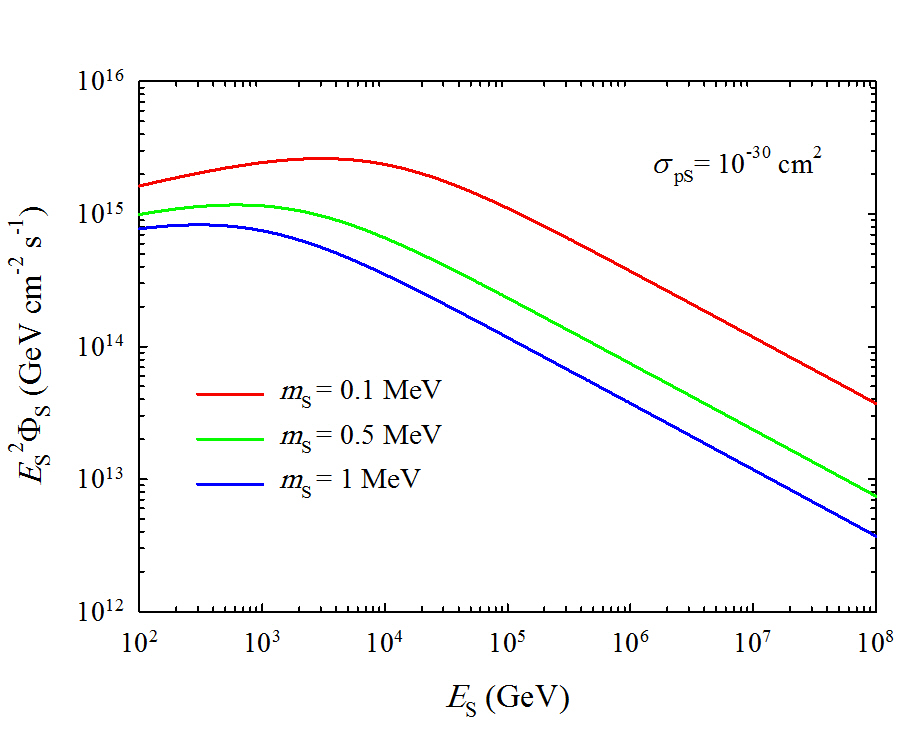}}
\caption{(Log-scale) DM energy flux resulting from interactions with extragalactic CR protons. In this Fig., we fix $m_S=0.1, 0.5$ and $1$ MeV. The DM-proton cross section is also set to $\sigma_{pS}=10^{-30} \text{cm}^2$. }
\label{fig.DMflx1}
\end{figure}

In Fig.~\ref{fig.DMflx1}, we show the energy spectrum of singlet scalar $S$ up-scattered by extragalactic protons. In this Fig., we indeed investigate the boosting of DM particles using Eq. \ref{Eq.14}. Here, the scattering cross section of DM with proton is assumed $\sigma_{pS}=10^{-30} \text{cm}^2$ and an statistical analysis is illustrated for three fixed DM mass $m_S=0.1, 0.5$ and $1$ MeV. The break occurs at $E_S\sim 2m_S/(m_p^2+m_S^2)$ for all three masses. At lower energies, prior to this spectral break, the extragalactic DM flux increases gradually and then shows a noticeable drop for HE DM. Furthermore, it is evident that a lighter DM particle corresponds to a more noticeable flux.

Compared to previous estimates such as Ref.~\cite{Guo:2020drq}, the DM flux shown in Fig.~\ref{fig.DMflx1} is enhanced due to the assumed local abundance of UHECRs produced by GRB 221009A. The dense environment near the GRB can significantly boost the scattering rate and subsequent DM flux. We note that although the resulting DM flux can be substantial, it remains within plausible astrophysical limits considering uncertainties in UHECR propagation and GRB energetics. In this context, the DM flux does not exceed the UHECR flux itself but reflects a transient enhancement due to GRB activity.

\begin{figure}[H]
	\centerline {\includegraphics[width=.60\textwidth]{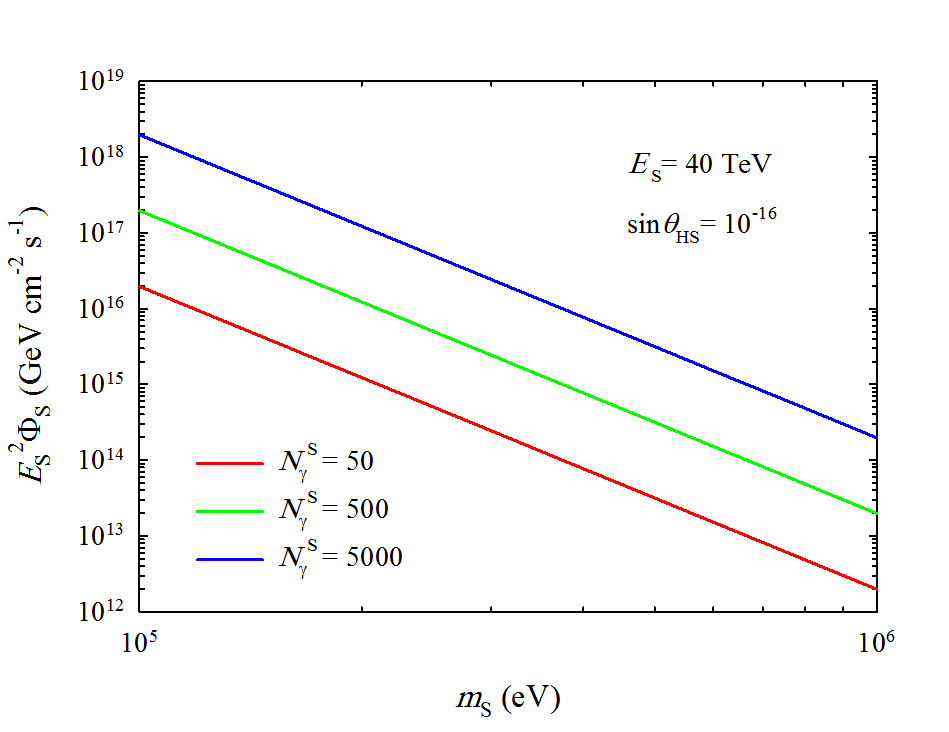} }
	\small\caption{(Log-scale) The DM energy spectrum as a function of DM mass $m_S$ for cases of 50, 500 and 5000 number of events. The energy of DM candidate $S$ is taken to be $E_S=40$ TeV and its mixing with the SM is parameterized as $\sin \theta_{HS}=10^{-16}$.}
	\label{fig.DMflx2}
\end{figure}

In a complementary analysis, using Eq. \ref{Eq.27}, we evaluate the role of the number of events in the incident DM flux at Earth. In this manner, the kinematics associated with two-body DM decay, along with the relic density constraint, yield values of $E_S=40$ TeV and $\sin \theta_{HS}=10^{-16}$. The effect is prominent for three obtained values $N_\gamma^S=50,500$ and 5000 as depicted in Fig.~\ref{fig.DMflx2}. As shown in this figure, the DM flux exhibits an inverse relationship with $m_S$, consistent with the findings of the prior analysis, while demonstrating a direct proportionality to $N_\gamma^S$.    

The combination of Eqs. \ref{Eq.14} and \ref{Eq.27} yields the optimal fit for our model. In this sense, for DM mass $m_S\simeq1$ MeV, DM-proton cross section $\sigma_{pS}=10^{-30} \text{cm}^2$ and DM number density associated with a redshift of $z=0.1505$, we obtain a flux of $\Phi_S\simeq10^5$ GeV$^{-1}$ cm$^{-2}$ s$^{-1}$ (Eq. \ref{Eq.14}). Moreover, this flux corresponds to 5000 events produced from DM particles of the same mass as well as mixing $\sin \theta_{HS}=10^{-16}$ in Eq. \ref{Eq.27}. Thus, our framework offers a novel DM interpretation for GRB 221009A. It should be noted here that the assumed value of the up-scattering cross section $\sigma_{pS}$ is consistent with effective values obtained in certain simplified DM interaction models. Resonance effects or threshold enhancements may account for the relatively high cross section, without violating theoretical or observational constraints. A detailed exploration of such effects is left for future work.

\begin{figure}[H]
	\begin{subfigure}{0.5\textwidth}
		\includegraphics[width=1\linewidth, height=6.5cm]{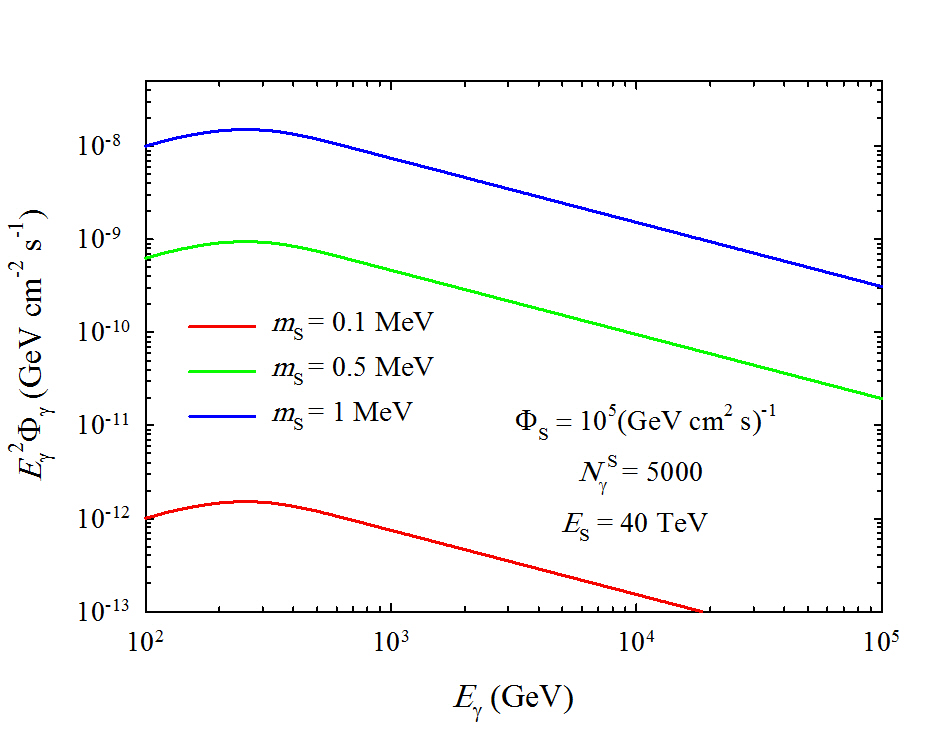} 
		\caption{}
		\label{fig.GRflx1}
	\end{subfigure}
	\begin{subfigure}{0.5\textwidth}
		\includegraphics[width=1\linewidth, height=6.5cm]{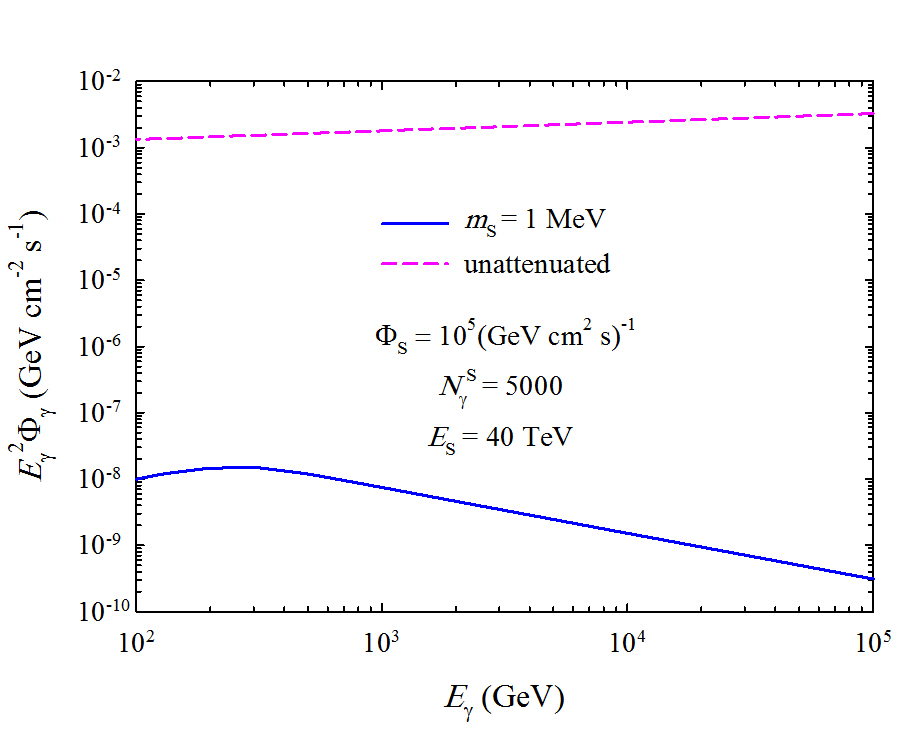}
		\caption{}
		\label{fig.GRflx2}
	\end{subfigure}
	\caption{(Log-scale) The emitted spectrum of GRB 221009A as a function of the photon energy $E_\gamma$. In this Fig., we set  $\Phi_S=10^5$ GeV$^{-1}$ cm$^{-2}$ s$^{-1}$, $N_\gamma^S=5000$ and $E_S= 40~ \text{TeV}$. (a) For  different values of DM mass $m_S$. (b) The highest resulting flux from DM decay, corresponded to $m_S=1$ MeV, is compared with the unattenuated photon flux.}
	\label{fig.GRflx}
\end{figure}

We now turn our attention to the $\gamma$ ray flux $\Phi_\gamma$ from the radiative decay diphoton DM$\rightarrow\gamma\gamma$. Substituting the aforementioned DM flux as well as other relevant factors in Eq. \ref{Eq.26}, we can compute spectrum of the most number of events by LHAASO. The result is shown in Fig.~\ref{fig.GRflx} in terms of gamma ray energy where parameters are taken as $\Phi_S=10^5$ GeV$^{-1}$ cm$^{-2}$ s$^{-1}$, $N_\gamma^S=5000$ and $E_S= 40~ \text{TeV} (>2E_\gamma)$.
 In panel (a), we examine three cases $m_S=0.1, 0.5$ and $1$ MeV so that we can argue the effect of DM mass on GRB flux at Earth. As illustrated in Fig.~\ref{fig.GRflx1}, lighter DM candidates generate very weak flux such that for $m_S\lesssim0.1$ MeV, $\gamma$ flux is suppressed (in contrast to the case for the DM flux (see Figs. \ref{fig.DMflx1} and ~\ref{fig.DMflx2})). Besides, it clearly shows that gamma ray flux decreases at larger energy regimes. In panel (b), we compare the most prominent flux (corresponding to $m_S=1$ MeV) with the case where all GRB's photons can reach Earth, dubbed as unattenuated gamma flux. This flux is extracted from Fermi-LAT data and is formulated as (\cite{Smirnov:2022suv})
 \begin{equation}
 \Phi_\gamma^0(E_\gamma)=\frac{2.1\times10^{-6}}{\text{cm}^2\text{sGeV}}(\frac{E_\gamma}{\text{GeV}})^{-1.87\pm 0.04}.
 \end{equation}    
It is clear from the latter analysis that $\gamma$ flux from DM decay can not approach the unattenuated flux.

We now demonstrate that how singlet scalar DM can acquire constraints from GRB 221009A. First, we considered DM with masses low enough ($m_S<2m_e$) to only decay to two photons. Second, the process of boosting DM indicated that light DM ($m_S\leq 1$ MeV) can gain enough kinetic energy and generate large flux related to the number of events expected in 2000 s. Finally, DM with $m_S=1$ MeV generated the most promising photon flux at Earth after its radiative decay. As a consequence, singlet scalar DM should have the mass $m_S=1$ MeV and mixing $\sin \theta_{HS}=10^{-16}$ (coupling strength $\lambda_{HS}=10^{-9}$) to fit LHAASO data of GRB 221009A measurement.








\section{Conclusion}\label{sec:con}
Due to the exceptionally energetic photon emission observed on 9th October, 2022, we proposed a novel hypothesis to elucidate this event. The choice of redshift \( z = 0.1505 \) is motivated by the unique characteristics of GRB 221009A. Its extreme brightness, long duration, and unprecedented photon energies make it a prime candidate for exploring exotic particle physics. Although other GRBs exist at comparable redshifts, none exhibit such an energetic profile. We emphasize that the detection of HE photons from GRB 221009A is a rare phenomenon, and the proposed mechanism may not operate efficiently in less extreme cases.

To achieve this, we employed BSM physics to mitigate the substantial cosmic suppression experienced by photons originating from outside the Milky Way galaxy. Thus, we adopted a singlet scalar field residing BSM and interacting with SM particles through Higgs boson $h$. DM particles travel a long distance from redshift $z=0.1505$ and undergo inelastic collisions with cosmic proton targets on the way to Earth. The scattering gives rise large fluxes of incident DM at Earth. This HE DM now is able to generate anomalous number of $\cal{O}$(5000) photons with $E_\gamma$ in the [0.5, 13] TeV range via DM$\rightarrow\gamma\gamma$. Terrestrial observatories, such as LHAASO \cite{2022GCN.32677....1H}, are equipped to detect the resultant HE gamma rays across this extensive energy spectrum.

The model we examined effectively parameterized GRB 221009A and was compatible with the observed $\gamma$ spectra. Following that, our scenario proposed DM decay as a novel explanation for GRB 221009A, addressing the challenges posed by photon attenuation in the intergalactic medium. In light of this phenomenological consideration, our DM model got new constraints within its parameter space based on LHAASO data. Consequently, the  best-fit for a singlet scalar DM was achieved at a mass of $m_S=1$ MeV and a mixing of $\sin \theta_{HS}=10^{-16}$. We have not only addressed a cosmological anomaly but have also established a novel avenue for the indirect detection of DM. Specifically, the rarity of GRB 221009A and its alignment with LHAASO observations underscore the uniqueness of this event, distinguishing it as an exceptional candidate for studying DM signatures.

\section*{ACKNOWLEDGMENTS}
We thank S Y Ayazi for providing useful comments to this manuscript. We also thank Y Farzan and Sh Balaji for helpful discussions.         





\end{document}